\documentclass[3p,times]{elsarticle}

\usepackage{ecrc}

\volume{00}

\firstpage{1}

\journalname{Physica A}

\runauth{M.~N.~\c{C}ankaya and J.~Korbel}

\jnltitlelogo{Physica A}

\usepackage{amsmath}
\usepackage{amssymb}
\usepackage{amsthm}
\usepackage{dsfont}
\usepackage{graphicx,color}
\newcommand{\ud}{\mathrm{d}}
\dochead{}
\begin{document}
\begin{frontmatter}

 %%%%%%%%%%%%%%%%%%%%%%%%%%%%%%%%%%%%%%%%%%%%%%%%%%%%%%%%%%%%%%%%%%%%%%%%%%%%%%%%%%%%%%%%%%%%%%%%%%
%\title{On Statistical Properties of Hybrid Entropy}% Force line breaks with \\
\title{On Statistical Properties of Jizba-Arimitsu Hybrid Entropy}% Force line breaks with \\
%%%%%%%%%%%%%%%%%%%%%%%%%%%%%%%%%%%%%%%%%%%%%%%%%%%%%%%%%%%%%%%%%%%%%%%%%%%%%%%%%%%%%%%%%%%%%%%%%%%%%
%
\author[USAK]{Mehmet Niyazi \c{C}ankaya}
\ead{mehmet.cankaya@usak.edu.tr}
\author[ZJU,CTU]{Jan Korbel}%
\ead{korbeja2@fjfi.cvut.cz}

\address[USAK]{U\c{s}ak University, Faculty of Art and Sciences, Department of Statistics, U\c{s}ak, Turkey}
\address[ZJU]{Department of Physics, Zhejiang University, Hangzhou 310027, P. R. China}
\address[CTU]{Faculty of Nuclear Sciences and Physical Engineering, Czech Technical University in Prague,
B\v{r}ehov\'{a} 7, 115 19 Praha 1, Czech Republic}

\begin{abstract}
Jizba-Arimitsu entropy (also called \emph{hybrid entropy}) combines axiomatics of R\'{e}nyi and Tsallis entropy. It has many common properties with them, on the other hand, some aspects as e.g., MaxEnt distributions, are completely different from the former two entropies. In this paper, we demonstrate the statistical properties of hybrid entropy, including the definition of hybrid entropy for continuous distributions, its relation to discrete entropy and calculation of hybrid entropy for some well-known distributions. Additionally, definition of hybrid divergence and its connection to Fisher metric is also discussed. Interestingly, the main properties of continuous hybrid entropy and hybrid divergence are completely different from measures based on R\'{e}nyi and Tsallis entropy. This motivates us to introduce average hybrid entropy, which can be understood as an average between Tsallis and R\'{e}nyi entropy
\end{abstract}
%\textbf{Highlights}

%\begin{itemize}
 % \item We present Hybrid entropy ($HE$) proposed by  \citep{Korbel2}.
  %\item We get the Hybrid entropy for the $\varepsilon$-skew exponential power distribution family and beta distribution.
%\end{itemize}

\end{frontmatter}

%%%%%%%%%%%%%%%%%%%%%%%%%%%%%%%%%%%%%%%%%%%%%%%%%%%%%%%%%%%%%%%%%%%%%%%%
\section{Introduction}
%%%%%%%%%%%%%%%%%%%%%%%%%%%%%%%%%%%%%%%%%%%%%%%%%%%%%%%%%%%%%%%%%%%%%%%%
Generalized entropies have played an important role in description of thermodynamic, statistical and informational systems in past few decades. The main reason for using these entropies is to describe systems that cannot be successfully described by the conventional Shannon-Boltzmann-Gibbs entropy. In information theory appeared Shannon entropy firstly in 1948 \cite{Shannon} in connection with communication theory. Shortly afterwards, there began to appear various generalizations of Shannon entropy. To the most important belong R\'{e}nyi entropy \cite{Renyi}, Tsallis(-Havrda-Charv\'{a}t) entropy (derived independently by Tsallis \cite{Tsallis} from thermodynamical point of view and by Havrda and Charv\'{a}t~\cite{Havrda} from informational point of view), Sharma-Mittal entropy \cite{Sharma}, Frank-Daffertshofer entropy \cite{Frank} or Kapur measure \cite{Kapur}. Recently, there have been made several successful attempts in order to categorize the various entropy classes and their properties. Hanel and Thurner \cite{Thurner,Hanel} classified the entropies according to their asymptotic scaling, Tempesta \cite{Tempesta} studied the generalized entropies according to group properties, Bir\'{o} and Barnaf \cite{Biro} derived a new class of entropies from its interaction with heat reservoir. Ili\'{c} and Stankovi\'{c} \cite{Ilic} classified the pseudo-additive entropies by generalization of Khinchin axioms.

Among these entropies, the most prominent two classes are R\'{e}nyi entropy, also known from theory of multifractal systems~\cite{Jizba4}, and Tsallis entropy, describing the thermodynamics of non-extensive systems (e.g., systems in contact with finite heat bath~\cite{Ramshaw}). Jizba and Arimitsu~\cite{Jizba} suggested a new one-parametric class of entropies called \emph{Jizba-Arimitsu hybrid entropy}, which shares some properties of R\'{e}nyi and Tsallis entropy. Particularly, hybrid entropy is non-extensive and the conditional entropy is similarly to R\'{e}nyi defined in terms of generalized Kolmogorov-Nagumo mean~\cite{Kolmogorov,Nagumo}. Recently, hybrid entropy was discussed by several authors. Ili\'{c} and Stankovi\'{c} correctly pointed out that there is a mistake in the original derivation and corrected the axiomatic~\cite{Stankovic}. Jizba and Korbel \cite{Korbel3} calculated the error in the original definition and concluded that for thermodynamical systems with weak interactions is still possible to use hybrid entropy in its original form. Moreover, thermodynamic properties of hybrid entropy were extensively discussed in~\cite{Korbel2}.

The main aim of the paper is to present the statistical properties of hybrid entropy. These properties are often remarkably different from analogous results in the case of R\'{e}nyi and Tsallis entropy, which points to the fact that hybrid entropy can describe conceptually different systems than the former two entropies. We start with the relation to other generalized entropies. Subsequently, we define the continuous (also called differential) hybrid entropy and present its relation to the discrete hybrid entropy. Finally, we introduce hybrid divergence and study their statistical and informational properties, which motivates us into definition of average hybrid entropy. The rest of the paper is organized as follows: the next section is dedicated to revision of basic properties of hybrid entropy, relation to the maximality axiom, the concavity issue and the relation to the R\'{e}nyi and Tsallis entropies. Section 3 defines the continuous hybrid entropy and shows its properties on popular distributions, such as triangular, Beta, exponential, Gamma, normal, Cauchy and Student-t distribution. In Section 4 is introduced hybrid divergence and and its informational properties are studies, especially its connection to Fisher metric. Section 5 defines average hybrid entropy and briefly discusses its properties. The last section is devoted to conclusions. Appendix presents the hybrid entropy for special class of bimodal $\varepsilon$-skew exponential power distributions, which are important in the theory of statistical inference.

%%%%%%%%%%%%%%%%%%%%%%%%%%%%%%%%%%%%%%%%%%%%%%%%%%%%%%%%%%%%%%%%%%%%%%%%
\section{Basic properties of hybrid entropy}
%%%%%%%%%%%%%%%%%%%%%%%%%%%%%%%%%%%%%%%%%%%%%%%%%%%%%%%%%%%%%%%%%%%%%%%%
Hybrid entropy is defined as a synthesis between Tsallis and R\'{e}nyi entropy and it combines both $q$-non-extensivity and generalized Kolmogorov-Nagumo $q$-averaging, which is important for multifractal systems, especially. The resulting hybrid entropy $\mathcal{D}_q(P)$ is defined by four axioms~\cite{Jizba}. The continuity axiom requires the continuity in every argument, the maximality axiom requests that the entropy is maximal for uniform distribution and expansibility axiom ensures that an event with zero probability does not affect the value of entropy. The most important additivity axiom defines the joint entropy and conditional entropy as

\begin{equation}
\mathcal{D}_q(A \cup B) = \mathcal{D}_q(A) + \mathcal{D}_q(B|A) + (1-q) \mathcal{D}_q(A) \mathcal{D}_q(B|A)
\end{equation}
resp.
\begin{equation}\label{eq: con}
\mathcal{D}_q(B|A) = f_q^{-1} \left( \sum_k \rho_k(q) f_q (\mathcal{D}_q(B|A = A_k)) \right)
\end{equation}
where $f_q$ is a positive, invertible function on $\mathds{R}^+$. Its exact form is discussed in Section \ref{sec: q}. Distribution $\rho_k(q) = p_k^q/\sum_j p_j^q$ is the \emph{escort distribution} belonging to experiment $A$. Escort distribution, or ``zooming distribution'', was originally discovered in connection with chaotic dynamic systems~\cite{Beck,Beck2} and is also widely used in theory of multifractals~\cite{Harte}.

It has been shown in Ref.~\cite{Stankovic} that the additivity rule holds exactly only for independent events. In~\cite{Korbel3} has been discussed the error for events which are not independent. This error is relatively small for weakly interacting systems. The form of the resulting hybrid entropy has been introduced in Ref.~\cite{Jizba} in the form:
\begin{equation}
\mathcal{D}_q(P) = \frac{1}{1-q} \left(e^{-(1-q) \sum_k \rho_k(q) \ln p_k} -1\right)\, .
\end{equation}

%-----------------------------------------------
\subsection{$q$-Deformed Calculus and Generalized Means}\label{sec: q}
%----------------------------------------------
In order to understand the properties of hybrid entropy, let us introduce two mathematical terms which are important in theory of generalized entropies. First, let us briefly revise so-called $q$-deformed calculus. This calculus is connected with non-extensive entropies, especially Tsallis entropy. The aim is to define the non-linear generalizations of ordinary operations and functions. For example, \emph{$q$-deformed addition} (also known as \emph{Jackson sum}~\cite{Qdef}) is defined as
\begin{equation}
x \oplus_q y = x+y+(1-q)xy\, .
\end{equation}
We can recognize that the additivity rule for hybrid entropy is nothing else than $q$-deformed sum. Additionally, we can define the $q$-deformed versions of logarithm and exponential functions~\cite{Box}:
\begin{eqnarray}
\exp_q(x \oplus_q y) = \exp_q(x) \exp_q(y) &\Rightarrow& \exp_q(x) = [1+(1-q)x]^{1/(1-q)}\\
\ln_q(xy) = \ln_q(x) \oplus_q \ln_q(y) &\Rightarrow& \ln_q(x) = \frac{x^{1-q}-1}{1-q}
\end{eqnarray}
For appropriate values, it holds that
\begin{equation}
\exp_q (\ln_q (x)) = \ln_q (\exp_q (x)) = x.
\end{equation}

Second, let us define two classes of generalized means. The first class is known as \emph{Kolmogorov-Nagumo means}~\cite{Kolmogorov,Nagumo} and is defined as
\begin{equation}
E[X]_f = f^{-1} \left(\sum_k p_k f(x_k) \right)
\end{equation}
Additionally, \emph{escort means}~\cite{Bercher} are based on escort distributions and are defined as
  \begin{equation}
E[X]^q = \left(\sum_k \rho_k(q) x_k \right)
\end{equation}
Of course, it is possible to combine both classes to obtain a generalized \emph{escort Kolmogorov-Nagumo mean}, which reads
\begin{equation}
E[X]_f^q = f^{-1} \left(\sum_k \rho_k(q) f(x_k) \right)\, .
\end{equation}
These means often appear in connection with generalized entropies and one can recognize that generalized entropy is nothing else than a generalized mean of its elementary information (compare with Eq.~(\ref{eq: con}), as shown in the next section.
%Naturally, the conditional hybrid entropy is nothing else than escort Kolmogorov-Nagumo mean.

%======================================================================
\subsection{Relation to R\'{e}nyi and Tsallis entropy}
%======================================================================
%\begin{figure}[t]
%\begin{center}
%\includegraphics[width=12cm]{q0.5.eps}\\
%\includegraphics[width=12cm]{q2.eps}
%\end{center}
%\caption{Comparison of Hybrid entropy $\mathcal{D}_q$, Tsallis entropy $\mathcal{S}_q$, R\'{e}nyi entropy $\mathcal{R}_q$ and Shannon entropy $\mathcal{H}$ on distribution $P= (p,1-p)$ as a function of $p$ for $q=0.5,2$. Shannon entropy corresponds to $q \rightarrow 1$ for all remaining entropies.}
%\end{figure}

Definition of hybrid entropy was motivated as an overlap between R\'{e}nyi~\cite{Renyi} and Tsallis~\cite{Abe} axiomatic and therefore it shares many properties with both of them. Let us first remind definitions of R\'{e}nyi entropy and Tsallis entropy:
\begin{eqnarray}
\mathcal{R}_q(P) &=& \frac{1}{1-q}\ln \sum_k p_k^q\\
\mathcal{S}_q(P) &=& \frac{1}{1-q} \left(\sum_k p_k^q -1\right)\,
\end{eqnarray}
where both are generalizations of Shannon entropy
\begin{eqnarray}
\mathcal{H}(P) &=& - \sum_k p_k \ln p_k\, .
\end{eqnarray}
All entropies become Shannon entropy for $q \rightarrow 1$. It is easy to see that these two entropies are functions of each other, for instance
\begin{eqnarray}\label{eq: tsaren}
\mathcal{R}_q(P) = \frac{1}{1-q} \ln [(1-q) \mathcal{S}_q(P) +1]\, .
\end{eqnarray}
This is not the case of hybrid entropy, because besides $\sum_k p_k^q$, there also appears the term $\sum_k p_k^q \ln p_k$. Connection of hybrid entropy to R\'{e}nyi entropy is given by the exponent $E[\ln P]^q$ which is equal to
\begin{equation}
E[\ln P]^q = \sum_k \rho_k(q) \ln p_k = \frac{\ud \left(\ln \sum_k p_k^q\right)}{\ud q} = (1-q) \frac{\ud \mathcal{R}_q(P)}{\ud q} - \mathcal{R}_q(P)\, .
\end{equation}
This term is widely known in the theory of multifractals and scaling, because it is closely related to multifractal spectrum and other scaling exponents~\cite{Korbel}. We can recognize that hybrid entropy is expressible in terms of R\'{e}nyi entropy and its derivative.

Let us focus on the contribution of elementary information to see the connection of hybrid entropy to $q$-additivity and Tsallis entropy. It can be derived that both entropies follow the $q$-additivity rule. Let us consider $m$ i.i.d. variables $\{A_i\}_{i=1}^m$ with the uniform distribution $p_k = \frac{1}{n}$. Plugging into additivity axiom, we obtain that
\begin{eqnarray}
\mathcal{D}_q\left(\frac{1}{n^m}\right) &=& \sum_{k=1}^m {m \choose k} (1-q)^{k-1} \mathcal{D}^k_q\left(\frac{1}{n}\right)\nonumber\\ &=& \frac{1}{1-q}\left[ \left(1+(1-q) \mathcal{D}_q\left(\frac{1}{n}\right)\right)^m -1 \right]\, .
\end{eqnarray}
Solution of this relation can be expressed as a $q$-deformed logarithm, i.e., $\ln_q(n)$. Thus, similarly to Tsallis entropy, the elementary information contribution, also called \emph{Hartley information}, of an event with probability $p_k$ is equal to
\begin{equation}
\mathcal{I}_q(p_k) = \ln_q\left(\frac{1}{p_k}\right).
\end{equation}
On the contrary, Hartley information of R\'{e}nyi entropy is (similarly to Shannon entropy) equal to
\begin{equation} \mathcal{I}_1(p_k) = \ln \left(\frac{1}{p_k}\right)\, ,
\end{equation}
 which is a consequence of additivity rule. It is clear that the elementary information can be determined from the additivity rule for independent events. Consequently, all introduced entropies can be represented as a generalized mean of the Hartley information~\cite{Hartley}. In the case of Tsallis entropy is the representation given by a simple mean of a non-extensive Hartley informations, i.e. as
\begin{equation}
\mathcal{S}_q(P) = E[\mathcal{I}_q(P)]
\end{equation}
while R\'{e}nyi entropy can be represented as a Kolmogorov-Nagumo mean of extensive Hartley information~\cite{Jizba2}. Alternatively, it can be viewed as an escort Kolmogorov-Nagumo mean with the exponential Kolmogorov-Nagumo function. This representation has been described in~\cite{Jizba2} as
\begin{equation}
\mathcal{R}_q(P) = E[\mathcal{I}_1(P)]_{\ln_q \exp x} = E[\mathcal{I}_1(P)]_{\exp[(q-1)x]}^q
\end{equation}
Finally, the hybrid entropy is expressible as a generalized Kolmogorov-Nagumo escort mean of its elementary information $\mathcal{I}_q(p_k)$
\begin{equation}
\mathcal{D}_q(P) = E[\mathcal{I}_q(P)]^q_{\ln \exp_q x}
\end{equation}
Note that Kolmogorov-Nagumo function of hybrid entropy is the inverse function R\'{e}nyi Kolmogorov-Nagumo function.

%=====================================================================
\subsection{Maximality axiom, concavity and Schur-concavity}
%=====================================================================
When deriving the entropies from Khinchin axioms, it is necessary to note that validity of maximality axiom (i.e., entropy is maximal for the uniform distribution) is not completely determined and therefore it is necessary to check its validity manually. This is generally quite complicated task. However, it is possible to investigate several properties which are sufficient to prove the validity. The most popular criterion is possibly concavity of entropy, because many entropy functionals, including Shannon and Tsallis, are concave functions. Moreover, if they belong to the \emph{trace class}~\cite{Hanel} defined as $s(P) = \sum_i g(p_i)$, concavity of entropy is equivalent to concavity of one-dimensional function $g(x)$. It can be shown that hybrid entropy is concave only in the interval $q \in [\frac{1}{2},1]$ (see Ref.~\cite{Korbel2}).

On the other hand, concavity is only sufficient property. It is possible to find weaker versions of concavity which also ensure the validity of maximality axiom. One of these concepts is so-called \emph{Schur-concavity}, extensively discussed e.g., in Ref.~\cite{Schur}, which is based on the theory of majorization. A discrete probability distribution $P = (p_1,\dots,p_n)$ is majorized by $Q = (q_1,\dots,q_n)$ if for their ordered probabilities $p_{(1)} \geq p_{(2)} \geq \dots \geq p_{(n)}$, resp. $q_{(1)} \geq q_{(2)} \geq \dots \geq q_{(n)}$ the following inequalities hold
\begin{equation}
\sum_{k=1}^j p_{(k)} \leq \sum_{k=1}^j q_{(k)}\, .
\end{equation}
We denote it as $P \prec Q$. A function $F$ is \emph{Schur-concave} if for every $P \prec Q$ is $F(P) \geq F(Q)$ (Analogously, $G$ is Schur-convex if for every $P \prec Q$ is $G(P) \leq G(Q)$). It is easy to show that the uniform distribution is majorized by any other distribution
\begin{equation}
\left(\frac{1}{n},\dots,\frac{1}{n}\right) \prec (p_1,\dots,p_n) \quad \forall P = (p_1,\dots,p_n) \quad s.t. \ \sum_i p_i = 1,
\end{equation}
so it is clear that Schur-concavity of every entropy functional is \emph{sufficient} property for validity of maximality axiom. It is weaker than concavity, which means that every symmetric concave function is Schur-concave. In \cite{Shi}, it was shown (as a special case of Schur-concavity of so-called Gini means) that hybrid entropy is Schur-concave for $q \geq \frac{1}{2}$, while for $q \in [0,\frac{1}{2})$ is neither Schur-convex nor Schur-concave, and it can be shown that it does not obey the maximality axiom.

%======================================================================
\subsection{MaxEnt distribution}
%======================================================================
MaxEnt distribution, originally proposed by Jaynes~\cite{Jaynes}, represents the distribution containing minimal amount of information under certain constraints. We always require the normalization of the probability distribution. Additionally, there are many other possible constraints. The most common is to prescribe an average energy $\mathcal{E}$, which is usually considered as an escort mean
\begin{equation}
\mathcal{E}= E[\mathcal{E}]^r = \sum_k \rho_k(r) \mathcal{E}_k\, .
\end{equation}
Two most common choices are \emph{linear averaging}, i.e., $r=1$ ($\sum_k p_k \mathcal{E}_k$) and $q$-averaging, i.e., $r=q$. The reason is that only these two cases provide a unique real MaxEnt distribution~\cite{Zatloukal}. Thus, maximization under constraints is equal to maximization of \emph{Lagrange function} which reads:
\begin{equation}
\mathcal{L}(P) = \mathcal{D}_q(P) - \alpha \left(\sum_k p_k\right) - \beta \left(\sum_k \rho_k(r) \mathcal{E}_k\right)\, .
\end{equation}
The MaxEnt distribution can be found by solving equations $\frac{\partial \mathcal{L}(P)}{\partial p_i} = 0$, which is equal to
\begin{eqnarray}\label{eq: max}
\frac{\partial \mathcal{L}(P)}{\partial p_i} = e^{(q-1) E[ln P]^q} \left\{ q \left(E[\ln P]^q - \ln p_i\right) - 1 \right\} \frac{p_i^{q-1}}{\sum_k p_k^q} -\nonumber\\
\alpha p_i - \beta r(\mathcal{E}_r - \langle\mathcal{E}\rangle_r) \frac{p_i^r}{\sum_k p_k^r} = 0\, .
\end{eqnarray}
Multiplying by $p_i$ and summing over $i$, we obtain that
\begin{equation}
\alpha = - e^{(q-1) E[\ln P]^q}\, .
\end{equation}
Plugging back into Eq. (\ref{eq: max}), we obtain the equation for $p_i$:
\begin{equation}
\alpha \left[\left\{ q \left(E[\ln P]^q - \ln p_i\right) - 1 \right\} \frac{p_i^{q-1}}{\sum_k p_k^q}+1\right] + r \beta (\mathcal{E}_r - \langle \mathcal{E} \rangle_r)\frac{p_i^r}{\sum_k p_k^r} = 0\, .
\end{equation}
The equation is intractable, except for the two aforementioned cases, i.e., $r=1$ and $r=q$. For these two cases, it is possible to express the solution in terms of \emph{Lambert $W$-function}
\begin{equation}
p_i =\left\{
       \begin{array}{ll}
         \left[\frac{1}{z_q} W\left(z_q e^{\frac{q-1}{q}(1-\frac{q \ln (-\alpha)}{q-1}- \frac{q \beta}{\alpha} \Delta_q \mathcal{E}_i}\right)\right]^{1/(1-q)} & \hbox{for r=q,} \\
         \left[ \frac{ \alpha}{z_q(\alpha + \beta {\Delta_1 \mathcal{E}_i})} W \left(\frac{z_q}{\alpha} \exp\left(\frac{q}{q-1}\right)\left\{1+\frac{\beta}{\alpha} \Delta_1 \mathcal{E}_i\right\}\right)\right]^{1/(1-q)} & \hbox{for r=1}
       \end{array}
     \right.
\end{equation}
where $z_q = \frac{(q-1) \sum_k p_k^q}{q}$ and  $\Delta_r \mathcal{E}_i = \mathcal{E}_i - \langle \mathcal{E} \rangle_r$. Lambert $W$-function is a solution of equation
\begin{equation}
x = W(x) e^{W(x)}\, .
\end{equation}
More details about Lambert function can be found e.g., in Ref.~\cite{Corless}. Particularly interestings are three cases: a) systems described by multifractal scaling exponents, b) ``high-temperature limit'' ($\beta \ll 1$), c)``low-temperature limit'' ($\beta \gg 1$). These three cases provide the interesting examples of complex dynamics of systems driven by the hybrid entropy and are extensively discussed in Ref.~\cite{Korbel2}.

%======================================================================
\section{Continuous hybrid entropy}
%======================================================================
In this section, we focus on definition of hybrid entropy for continuous distributions and present some connections to discrete hybrid entropy. Finally, we calculate hybrid entropy for some popular distributions and compare it with other entropies, mainly Tsallis entropy. Similarly to other cases~\cite{Jizba3}, continuous hybrid entropy can be defined as
\begin{equation}\label{AnalHybrid}
  \mathcal{D}_q[p(x)] = \frac{1}{1-q}\bigg[exp\bigg(- \frac{\int_\mathbb{S} p^q(x) log[p(x)] \ud x}{\int_\mathbb{S} p^q(x) dx}  \bigg)^{1-q}  - 1   \bigg]
\end{equation}
where $\mathbb{S}$ denotes support of probability distribution $p(x)$. First of all, it is necessary to point out that some properties of discrete entropies do not hold for continuous entropies. For example, the positivity of entropy functionals is not guaranteed for continuous distributions. This is also the case of continuous hybrid entropy.

Second, when establishing connection between the discrete entropy and its continuous analogue, the most common way is to think about the continuous entropy as a limit of discrete entropy for $n \rightarrow \infty$. Nevertheless, this limit is not convergent and it is necessary to make a renormalization. Let us consider a finite support $\mathbb{S} = [0,1]$. Then, it is possible to define a discrete approximation of $p(x)$ as
\begin{equation}
p_{k}^{(n)} = \int_{(k-1)/n}^{k/n} p(x) \, \ud x\,
\end{equation}
for $k \in \{1,\dots,n\}$, which is nothing than a histogram. Naturally, $p^{(n)}$ approximates $p(x) \ud x$ for large values of $n$. Thus, when $n \gg 1$ we can approximate $p(x)$ for $x \in [(k-1)/n,k/n]$ as $n p_{k}^{(n)}$. Consequently, the continuous hybrid entropy can be approximated as (here we omit the dependence of $p_k^{(n)}$ on $n$)
\begin{eqnarray}
\mathcal{D}_q[p(x)] = \ln_q \exp\left(- \frac{\sum_{k=1}^n \int_{(k-1)/n}^{k/n} p^q(x) \ln[p(x)] \, \ud x}{\sum_{k=1}^n \int_{(k-1)/n}^{k/n} p^q(x) \, \ud x}\right) \nonumber\\
\approx \ln_q \exp\left(- \frac{\sum_{k=1}^n \int_{(k-1)/n}^{k/n} \left(n p_k\right)^q \ln\left[n p_k\right] \, \ud x}{\sum_{k=1}^n \int_{(k-1)/n}^{k/n} \left(n p_k\right)^q  \, \ud x}\right)\nonumber \\ = \ln_q \exp\left( - \frac{n^{q-1} \sum_{k=1}^n p_k^q \ln p_k + n^{q-1}\ln n \sum_{k=1}^n p_k^q}{n^{q-1} \sum_{k=1}^n p_k^q}\right) = \nonumber\\
= \ln_q \exp\left(-\frac{\sum_{k=1}^n p_k^q \ln p_k}{\sum_{k=1}^n p_k^q}\right) \oplus_q \ln_q \frac{1}{n}\, .
\end{eqnarray}
Therefore, in order to process the limit $n \rightarrow \infty$, we have to renormalize $\mathcal{D}_q(p^{(n)})$ in order to keep the whole expression finite. Consequently, it is possible to express the relation between discrete and continuous hybrid entropy as
\begin{equation}\label{eq:condis}
\mathcal{D}_q[p(x)] = \lim_{n \rightarrow \infty} \left(\mathcal{D}_q(p^{(n)}) \oplus_q \ln_q \frac{1}{n}\right)\,
\end{equation}
if the limit exists. The situation is analogous if $\mathbb{S}$ is any other interval. It is necessary to mention that the renormalization in Eq.~(\ref{eq:condis}) is different from renormalization of R\'{e}nyi and Tsallis entropy. These can be found e.g., in~\cite{Jizba3}.

In the rest of this section, we compare continuous entropies for several popular distributions. Table~\ref{tab} compares Tsallis and hybrid entropy for several popular distributions. These include distributions with the finite support, the positive real support and the real support.  R\'{e}nyi entropy can be easily deduced from Tsallis by relation~(\ref{eq: tsaren}). Moreover, in Appendix, we also derive a hybrid entropy for $\varepsilon$-skew exponential power distribution~\cite{Elsalloukh}, which is a special class of bimodal distributions, recently introduced by \c{C}ankaya et al~\cite{Cankaya}. This class of distribution finds its place in the mathematical and physical problems, especially in statistical estimation. We can immediately recognize different dependence on parameter $q$ for Tsallis and hybrid entropy. Naturally, the functional dependence is also different, including more advanced classes of special functions, including Harmonic numbers, digamma function, etc. This is caused the fact that the hybrid entropy is conceptually different from Tsallis, which becomes even more evident in the next section, when the hybrid divergence is defined.

\begin{table}[t]
\begin{center}
\begin{tabular}{|c||c|c|}
  \hline
  Distribution
%& $\mathcal{R}_q(p)$
& $\mathcal{S}_q(p)$ & $\mathcal{D}_q(p)$ \\
  \hline
  \hline
%\begin{tabular}{c}Finite support\\
%$ \mathbb{S} = [0,1]$
%\end{tabular}& &\\
\multicolumn{3}{c}{Finite support: $ \mathbb{S} = [0,1]$}  \\ \hline
\hline
%%%%%%%%%%%%%%%%%%%%%%
\begin{tabular}{c}
Triangular:\\
\scriptsize $4x$ for $x\leq \frac{1}{2}$\\ \scriptsize $4(1-x)$ for $x\geq \frac{1}{2}$
 \end{tabular}
%& $\frac{\ln \left(\frac{2^q}{1+q}\right)}{1-q}$
& $\frac{\frac{2^q}{1+q}-1}{1-q}$
&$  \frac{\left( 2e^\frac{1}{q+1}\right)^{1-q}-1}{1-q}$\\
\hline
%%%%%%%%%%%%%%%%%%%%%%%%
\begin{tabular}{c}
Beta:\\
$\frac{x^{\alpha-1}(1-x)^{\beta-1}}{B(\alpha,\beta)}$
\end{tabular}
& \scriptsize $\frac{\frac{\Gamma (q (\alpha -1)+1) \Gamma (q (\beta -1)+1) \left(\frac{1}{B(\alpha ,\beta )}\right)^q}{\Gamma (q (\alpha +\beta -2)+2)}-1}{1-q}$
& %\tiny $\frac{1}{1-q}exp\bigg( \frac{\{(\alpha-1)^2+(\beta-1)\}[\psi(\beta q - q + 1) - \psi(\beta q + \alpha q - 2q + 2)] \frac{B(\beta q-q+1,\alpha q - q + 1)}{B^q(\alpha,\beta)}-\frac{B^q(\alpha,\beta)}{log(B(\alpha,\beta))}B(\alpha q - q +1, \beta q  - q +1)}{-B^q(\alpha,\beta)B(\alpha q - q +1, \beta q  - q +1)} \bigg)-1$
$\frac{\left(B(\alpha,\beta )  \exp\left[(\alpha +\beta -2) H_{q (\alpha +\beta -2)+1}+(1-\alpha ) H_{q (\alpha -1)}+(1 -\beta) H_{q (\beta -1)}\right]  \right)^{1-q}-1}{1-q}$
\\
\hline
\hline
%%%%%%%%%%%%%%%%%%%%%%%%%%
%\begin{tabular}{c}Half-plane\\
% $ \mathbb{S} = [0,\infty)$
%\end{tabular} & &\\
\multicolumn{3}{c}{Half-plane: $ \mathbb{S} = [0,\infty)$}  \\ \hline
\hline
%%%%%%%%%%%%%%%%%%%%%%%%%%%%%%%%%
     \begin{tabular}{c}Exponential: \\ $ \sigma \, e^{-\sigma x}$\end{tabular}
%& $\frac{\ln \left(\frac{\sigma^{q-1}}{q}\right)}{1-q}$
 & $\frac{\frac{\sigma^{q-1}}{q}-1}{1-q}$
& $\frac{\left(\frac{e^{1/q}}{\sigma}\right)^{1-q}-1}{1-q}$\\
\hline
%%%%%%%%%%%%%%%%%%%%%%%%%%%%%%%%%
\begin{tabular}{c}Gamma:\\
 $\frac{e^{- x/\beta} x^{\alpha-1} \beta^{-\alpha}}{\Gamma{(\alpha)}}$\end{tabular}
%&  $\frac{\log \left(\left(\frac{q}{\beta }\right)^{\alpha  q+q-1} \beta ^{\alpha  (-q)} \Gamma (\alpha )^{-q} \Gamma (1-(\alpha +1) q)\right)}{1-q}$
&  $\frac{\left(\frac{q}{\beta }\right)^{\alpha  q+q-1} \beta ^{\alpha  (-q)} \Gamma (\alpha )^{-q} \Gamma (1-(\alpha +1) q)-1}{1-q}$
& $\frac{\left( \left(\frac{q}{\beta }\right)^{1-\alpha } \exp\left[{\alpha +(\alpha -1) \psi  (1-(\alpha +1) q)-\frac{1}{q}+1} \left(\frac{\beta ^{-\alpha }}{\Gamma (\alpha )}\right)^{\Gamma (\alpha )^{q-1}}\right] \right)^{1-q}-1}{1-q}$ \\
%%%%%%%%%%%%%%%%%%%%%%%%%%%%%%%%%%
  %  &   &   &   \\
 \hline
 \hline
%%%%%%%%%%%%%%%%%%%%%%%%%%%%%%%%%
%\begin{tabular}{c}Whole plane\\
%$ \mathbb{S} = (-\infty,\infty)$
%\end{tabular} & &\\
\multicolumn{3}{c}{Whole plane: $ \mathbb{S} = (-\infty,\infty)$}  \\ \hline
\hline
%%%%%%%%%%%%%%%%%%%%%%%%%%%%%%%%%
   \begin{tabular}{c} Gaussian: \\
$\frac{1}{\sqrt{2 \pi \sigma^2}}\,  \exp(-\frac{(x-\mu)^2}{2\sigma^2})$ \end{tabular}
% &$\frac{\ln \left( \frac{(\sqrt{2 \pi \sigma^2})^{1-q}}{\sqrt{q}}\right)}{1-q}$
 & $\frac{ \frac{(\sqrt{2 \pi \sigma^2})^{1-q}}{\sqrt{q}}-1}{1-q}$
& $\frac{\left(\sqrt{2 \pi \sigma^2} \exp\left(\frac{1}{2q}\right) \right)^{1-q}-1}{1-q}$ \\
\hline
%%%%%%%%%%%%%%%%%%%%%%%%%%%%%%%%%
   \begin{tabular}{c} Cauchy: \\
$\frac{1}{\pi(1+x^2)}$ \end{tabular}
% &$\frac{\ln \left( \frac{(\sqrt{2 \pi \sigma^2})^{1-q}}{\sqrt{q}}\right)}{1-q}$
 & $\frac{\frac{\pi ^{\frac{1}{2}-q} \Gamma \left(q-\frac{1}{2}\right)}{\Gamma (q)}-1}{1-q}$
& $\frac{\exp\left[\,\frac{2^{q-1} \pi^{q-\frac{1}{2}} \Gamma \left(\frac{q+1}{2}\right)^2 \left(H_{\frac{q}{2}}-H_{\frac{q-1}{2}}+2 \log (\pi )\right)}{q \Gamma \left(q-\frac{1}{2}\right)}\right]^{1-q}-1}{1-q}$ \\
\hline
\begin{tabular}{c} Student: \\
$\frac{\Gamma((\nu+1)/2)}{\sqrt{\nu \pi}\Gamma(\nu/2)\sigma} \bigg[1+\frac{(x-\mu)^2}{\nu \sigma^2} \bigg]^{\frac{-\nu-1}{2}}$\end{tabular}
& $(\frac{\Gamma((\nu+1)/2)}{\sqrt{\nu \pi}\Gamma(\nu/2)})^q \sigma^{1-q}  \sqrt{\nu \pi} \frac{\Gamma(\frac{vq+q-1}{2})}{\Gamma(\frac{vq+q}{2})} $
&$\frac{\left(\frac{\Gamma((\nu+1)/2)}{\sqrt{\nu \pi}\Gamma(\nu/2)\sigma}\right)^{1-q} \exp\left[\frac{\sigma}{2} \frac{(\nu+1)}{ \pi^{1/2}} \bigg[\psi(\frac{ \nu q + q -1}{2})  -  \psi(\frac{ \nu q + q }{2})\bigg]\right]^{1-q} -1}{1-q}$\\

\hline
\end{tabular}
\caption{Comparison of continuous Tsallis entropy and hybrid entropy for several distributions. These include distributions with finite support (triangular, Beta), on the positive half-plane (exponential, Gamma) and on the whole real axis (Gaussian, Cauchy, Student). The resulting entropies are given in terms of elementary functions and certain special functions, including gamma function $\Gamma(z)$, beta function $B(\alpha,\beta)$ digamma function $\psi(z)$ and harmonic numbers $H_n$. Definitions and properties of these functions can be found, e.g. in Refs.~\cite{Ryzhik,Andrews}. }
\label{tab}
\end{center}
\end{table}

%Negentropy is a tool to measure departing off from the normality of a random variable $X$. It is defined as:
%\begin{equation}\label{negentropy}
%  H_{Neg}(X) = H(X_{Gauss}) - H(X)
%\end{equation}
%\noindent where $H$ shows the entropy of random variable $X$, $X_{Gauss}$ is a random variable for the Gaussian (normal) distribution. Negentropy is nonnegative, and will become larger when a random variable $X$ is far from the normality. Since $H$ is an entropy function, we can give the similar discussion for the hybrid entropy \citep{Hyvarinen01}.

%%%%%%%%%%%%%%%%%%%%%%%%%%%%%%%%%%%%%%%%%%%%%%%%%%%%%%%%%%%%%%%%%%%%%%%%
\section{Hybrid divergence and Fisher metric}\label{HybridDFisherM}
%%%%%%%%%%%%%%%%%%%%%%%%%%%%%%%%%%%%%%%%%%%%%%%%%%%%%%%%%%%%%%%%%%%%%%%%
Divergence, or relative entropy defines a relative distance of distribution $P$ from the underlying distribution $P_0$~\cite{Amari,Kullback}. While entropy is expressed as a functional of the probability density function $p = \frac{\ud P}{\ud \mu}$ on a measurable space with measure $\mu$, divergence is a functional of \emph{Radon-Nikodym derivative} $\frac{\ud P}{\ud P_0}$ on a space with measure $P_0$. Therefore, \emph{hybrid divergence} can be defined straightforwardly as
\begin{equation}
\mathcal{D}_q(P||P_0) = \mathcal{D}_q \left( \frac{\ud P}{\ud P_0} \right)_{P_0} = \ln_{q} \exp\left(- \int \rho_q \! \left( \frac{\ud P}{\ud P_0}\right) \, \ln \! \left(\frac{\ud P}{\ud P_0}\right) \ud P_0 \right)
\end{equation}
where $\rho_q\! \left( \frac{\ud P}{\ud P_0}\right)$ is the \emph{relative escort distribution}
\begin{equation}
\rho_q \! \left( \frac{\ud P}{\ud P_0}\right) = \frac{\left(\frac{\ud P}{\ud P_0}\right)^q}{\int \left( \frac{\ud P}{\ud P_0}\right)^q \ud P_0}\, .
\end{equation}
If there exist probability density functions $p$, resp. $p_0$, it is possible to rewrite the hybrid divergence as
\begin{equation}
\mathcal{D}_q(p||p_0) = \ln_{q} \exp\left(- \frac{\int p^q p_0^{1-q} \ln\left(\frac{p}{p_0}\right) \ud \mu}{\int p^q p_0^{1-q} \ud \mu}\right)
\end{equation}
Naturally, the hybrid divergence is well defined for $q \geq \frac{1}{2}$, while for $q < \frac{1}{2}$ it is not always positive for $p \neq p_0$.

Let us focus on a well-known connection between the entropies and the Riemann geometry on space of parametric probability distributions given by Fisher metric~\cite{Cover}. Let us consider a parametric family of distributions defined by parametric manifold $\mathcal{I}_\theta \subset \mathds{R}^n$:
\begin{equation}
\mathcal{F}_\theta = \{p({\bf x};\theta)|\,{\bf x} \in M, \theta \in \mathcal{I}_\theta\}\, .
\end{equation}
We define a pseudo-distance measure by symmetrization of hybrid divergence
\begin{equation}
d_q(p_1,p_2) = \frac{\mathcal{D}_q(p||p_0)+\mathcal{D}_q(p_0||p)}{2}\, .
\end{equation}
It is necessary to mention that $d_q(p_1,p_2)$ is not a proper distance measure, because it does not obey the triangle inequality. Nevertheless, for very close distributions it defines a metric tensor, because for $p_1({\bf x}) = p({\bf x},\theta)$ and $p_2({\bf x}) = p({\bf x},\theta + \ud \theta)$ we get that $d_q(p(\theta),p(\phi)) \equiv d_q(\theta,\phi)$ is equal to
\begin{eqnarray}
d_q(\theta,\theta+\ud \theta) &=&\frac{1}{2!}\sum_{i,j} \left[ \frac{\partial^2 d_q(\theta,\phi)}{\partial \theta_i \partial \theta_j}\right]_{\theta=\phi} \ud \theta_i \ud \theta_j + \mathcal{O}(\theta^3)\, \nonumber \\&=& \frac{1}{2!}\sum_{i,j} g_{ij}^q(\theta) \ud \theta_i \ud \theta_j + \mathcal{O}(\theta^3)\, .
\end{eqnarray}
Note that the absolute term vanishes, because $d_q(\theta,\theta) = 0$ and linear term vanishes, because $d_q(\theta,\phi)$ is minimal for $\theta=\phi$. The metric tensor is, for $q=1$, i.e., for Kullback-Leibler divergence, equal to $g_{ij}^1 = -  F_{ij}(\theta)$, where $F_{ij}$ is the Fisher information matrix
\begin{equation}
F_{ij}(\theta) = E\left[\frac{\partial \ln p(x,\theta)}{\partial \theta_i}\frac{\partial \ln p(x,\theta)}{\partial \theta_j}\right] = \int \frac{1}{p(x,\theta) }\frac{\partial p(x,\theta)}{\partial \theta_i}\frac{\partial p(x,\theta)}{\partial \theta_j} \, \ud x.
\end{equation}
After a straightforward calculation, we get that for hybrid divergence is the metric tensor equal to
\begin{equation}\label{generalizedFisher}
g_{ij}^q(\theta) =  (1-2q)\, F_{ij}(\theta) = (2q-1)\, g_{ij}^1(\theta) \, .
\end{equation}

It is necessary to note a few comments here. First, for $q = \frac{1}{2}$ is the metric identically equal to zero. This is caused by the fact the second derivative of $\mathcal{D}_{\frac{1}{2}}$ is equal to zero for $p = \{\frac{1}{n},\dots,\frac{1}{n}\}$. Thus, it is impossible to recognize two very close distributions from each other when $\mathcal{D}_{\frac{1}{2}}$ is used. Second, the metric is different from R\'{e}nyi, Tsallis and Sharma-Mittal metric of order $q$, because all are equal to $q g_{ij}^1(\theta)$~\cite{Bercher3,Bercher4}. The factor $(2q-1)$ is characteristic quantity for the whole hybrid entropy, obtained by combination of $q$-non-extensivity and $q$-escort averaging. Consequently, in the regime of close distributions, it is possible to relate the hybrid entropy parameter $q_\mathcal{D}$ with R\'{e}nyi entropy parameter $q_{\mathcal{R}}$, resp. Tsallis entropy parameter $q_{\mathcal{S}}$ as
\begin{equation}\label{eq: q}
q_{\mathcal{R}} = q_{\mathcal{S}} = 2 q_{\mathcal{D}}-1 \qquad \Rightarrow \qquad  q_{\mathcal{D}} = \frac{q_{\mathcal{R}}+1}{2}=\frac{q_{\mathcal{S}}+1}{2}\, .
\end{equation}
These conclusions motivate us to slightly modify the class of hybrid in order to obtain an average between Tsallis and R\'{e}nyi entropy.
%%%%%%%%%%%%%%%%%%%%%%%%%%%%%%%%%%%%%%%%%%%%%%%%%%%%%%%%%%%%%%%%%%%%%%%%
\section{Average hybrid entropy}
%%%%%%%%%%%%%%%%%%%%%%%%%%%%%%%%%%%%%%%%%%%%%%%%%%%%%%%%%%%%%%%%%%%%%%%%
Results of the last section motivate us to introduce a slightly different class of entropies, which would correspond to an average between R\'{e}nyi and Tsallis entropy, whereas hybrid entropy itself rather corresponds to combination of non-extensivity and escort averaging. Following Eq.~(\ref{eq: q}), we introduce \emph{average hybrid entropy} as
\begin{equation}
\mathcal{A}_q(p) = \mathcal{D}_{\frac{q+1}{2}}(p)\, .
\end{equation}

This entropy has many interesting properties. First, it is properly defined for $q > 0$, similarly to Tsallis and R\'{e}nyi. Moreover its Fisher metric is now the same as in the case of R\'{e}nyi and Tsallis. Naturally, for $q \rightarrow 1$, it boils down to Shannon entropy. Average hybrid entropy is concave for $q \leq 1$, while for $q > 1$ is only Schur-concave (similarly to R\'{e}nyi entropy).

Actually, the factor $\frac{q+1}{2}$ plays a special role in $q$-deformed calculus. One can easily see that
\begin{equation}
x \oplus_{\frac{q+1}{2}} y = x + y + \left(1- \frac{q+1}{2}\right) xy = x+ y + \frac{1-q}{2} xy = 2 \left(\frac{x}{2} \oplus_q \frac{y}{2}\right)\, .
\end{equation}
This means that the non-extensivity of average entropy can be expressed through non-extensivity of order $q$, but for rescaled quantities. Let us note that the non-extensivity parameter $\frac{q+1}{2}$ is actually an average between Tsallis non-extensivity parameter $q$ and R\'{e}nyi (non)-extensivity parameter $1$. Additionally, deformed logarithm appearing in the Kolmogorov-Nagumo function can be for average hybrid entropy expressed as
\begin{equation}
\ln_{\frac{q+1}{2}}(x) = \frac{x^{1-(q+1)/2}-1}{1-\frac{q+1}{2}} = 2 \frac{x^{\frac{1-q}{2}}-1}{1-q} = 2 \ln_q (\sqrt{x})\, .
\end{equation}
The deformed logarithm plays also role in the expression for maximal entropy. It is possible to show that
\begin{equation}
\ln_q n \geq \ln_{\frac{q+1}{2}} n \geq \ln n \quad \mathrm{for} \ q \leq 1, \qquad \qquad \ln_q n \leq \ln_{\frac{q+1}{2}} n \leq \ln n \quad \mathrm{for} \ q \geq 1\, .
\end{equation}
As a consequence, maximum of average hybrid is always between maxima of Tsallis and R\'{e}nyi entropy and can be considered as their average.
%%%%%%%%%%%%%%%%%%%%%%%%%%%%%%%%%%%%%%%%%%%%%%%%%%%%%%%%%%%%%%%%%%%%%%%%
\section{Conclusions}
%%%%%%%%%%%%%%%%%%%%%%%%%%%%%%%%%%%%%%%%%%%%%%%%%%%%%%%%%%%%%%%%%%%%%%%%
Hybrid entropy represents an conceptual overlap between Tsallis entropy and R\'{e}nyi entropy. Naturally, it shares many properties of both entropies: non-extensivity of Tsallis entropy and related upper bound, Shur-concavity, which appears at R\'{e}nyi entropy and much more. On the other hand, some properties are conceptually different, starting from accessible values of $q$ ($q \geq 1/2$), going through functional form of MaxEnt distribution, which can be expressed in terms of Lambert function (for Tsallis and R\'{e}nyi we obtain well-known $q$-deformed Gaussian distributions), and finally completely different properties of continuous hybrid entropy and hybrid divergence. The latter enables to establish connection to Fisher metric, which plays an important role in description of parametric probability distributions. Fisher metric obtained from hybrid divergence is conceptually different from R\'{e}nyi, Tsallis and Sharma-Mittal Fisher metric. As a result, we can identify the relation between hybrid entropy parameter $q_\mathcal{D}$ and Tsallis/R\'{e}nyi entropy parameter $q_\mathcal{S}$, resp. $q_\mathcal{R}$. As a result, it is possible to define the average hybrid entropy, which main properties can be interpreted as an average between Tsallis entropy and R\'{e}nyi entropy. All these results point to potential usefulness of (average) hybrid entropy in mathematical, physical or statistical applications. For a given entropy, it is possible to find a so-called Cram\'{e}r-Rao bound~\cite{Bercher4}, which connects the score function of a parametric distribution with inverse Fisher information obtained by the prescribed entropy functional. By utilization of hybrid entropy one can obtain a more precise bound for the estimations of model parameters. These possible applications are a subject of ongoing research.
%%%%%%%%%%%%%%%%%%%%%%%%%%%%%%%%%%%%%%%%%%%%%%%%%%%%%%%%%%%%%%%%%%%%%%%% If there is an instability, the data set should be modelled by the heavy-tailed distributions.In generally, the approach helps to classify the data set that are in the stable or unstable.
\section{Acknowledgements}
%%%%%%%%%%%%%%%%%%%%%%%%%%%%%%%%%%%%%%%%%%%%%%%%%%%%%%%%%%%%%%%%%%%%%%%%
We acknowledge helpful conversations with Petr Jizba. J.K. was supported by the GA\v{C}R, grant No. GA14-07983S.

\appendix

\section{Continuous hybrid entropy for $\varepsilon$-skew exponential power distribution}
In this appendix, we calculate the hybrid entropy for recently introduced $\varepsilon$-skew exponential power distribution (ESEP). The distribution was originally introduced in~\cite{Elsalloukh,Cankaya} as
\begin{equation}\label{specialBEGG}
  p(x;\mu,\sigma,\alpha,\beta,\eta,\varepsilon)=\frac{\alpha\beta}{2\sigma\eta^{1/\alpha}\Gamma(\frac{1}{\alpha\beta})}exp\{-\frac{|x - \mu|^{\alpha \beta}}{\eta^\beta(\sigma(1-sign(x-\mu)\varepsilon))^{\alpha\beta}} \}, \ \ x \in \mathds{R}
\end{equation}
where $\alpha,\beta,\eta,\sigma>0$, $\mu \in \mathbb{R}$ and $\varepsilon \in [-1,1]$. Parameters $\mu$ and $\sigma$ are location and scale parameters, respectively. Parameters $\alpha$ and $\beta$ are shape parameters that control the shape of peakedness. Parameter $\eta$ is a parameter describing the underlying kernel of normal or Laplace distribution. It can also be considered as a scale variant parameter. For $\alpha=2,\beta=1$ and $\eta=2$,  we get the $\varepsilon$-skew form for the normal distribution proposed by \cite{Mud}. ESEP is considered not only as a generalized version of normal distribution and its $\varepsilon$-skew form but also it is a class of the exponential power distributions. Since there are parameters that control the shape of density, it is likely to find more applications in statistics.

Let us have $\varepsilon$-skew exponential power distribution. In order to calculate hybrid entropy of ESEP distribution, we need to express following integrals:
\begin{eqnarray*}
  I_{ESEP_N} = \int_{-\infty}^\infty  p(x;\alpha,\beta,\eta,\varepsilon)^q \ln[ p(x;\alpha,\beta,\eta,\varepsilon)] \ud x \\
= \frac{2\eta^{(1/\alpha)(1-q)}\sigma^{1-q}(\alpha\beta)^{q-1}\Gamma(\frac{1}{\alpha\beta q})log(c)}{q [2\Gamma(\frac{1}{\alpha\beta})]^q}\\
-\bigg[\frac{(\alpha\beta)^{q-1}\eta^{\beta+1/\alpha(1-q)}\sigma^{\alpha\beta+1-q}}{q[2\Gamma(\frac{1}{\alpha\beta})]^q} \Gamma(\frac{1}{q}+\frac{1}{\alpha\beta q})  \bigg] \bigg[ (1+\varepsilon)^{\alpha\beta+1} + (1-\varepsilon)^{\alpha \beta + 1}  \bigg]
\end{eqnarray*}
where $c$ is the normalizing constant, $c=\frac{\alpha\beta}{2\sigma\eta^{1/\alpha}\Gamma(\frac{1}{\alpha\beta})}$. Similarly,

\begin{equation*}
 I_{ESEP_D} = \int_{-\infty}^\infty  p(x;\alpha,\beta,\eta,\varepsilon)^q \ud x= \frac{\bigg[2\sigma \eta^{1/\alpha}(\alpha\beta)^{-1}\bigg]^{1-q}}{\Gamma^q(\frac{1}{\alpha \beta})q}\Gamma(\frac{1}{\alpha\beta q}).
\end{equation*}

Finally, the hybrid entropy of ESEP distribution can be expressed with help of $I_{ESEP_N}$ and $I_{ESEP_D}$ as
\begin{equation}\label{propositionHEEPD}
\mathcal{D}_q =  \ln_q exp \left(- \frac{I_{ESEP_N}}{I_{ESEP_D}}  \right) = \ln_q \left( c \exp\left[\frac{(\eta\sigma)^{\beta}\sigma^{\alpha}\Gamma(\frac{1}{q}(1+\frac{1}{\alpha \beta}))[(1+\varepsilon)^{\alpha\beta+1}+(1-\varepsilon)^{\alpha\beta+1}]}{2\Gamma(\frac{1}{\alpha \beta q})}\right] \right)
%***\mathrm{can \ you \ simplify \ it?}****
\end{equation}

Hybrid entropy depends on all parameters in the $\varepsilon-$skew exponential power distribution, except the location parameter, which is similar to other distributions, see e.g.~\cite{KulMulti}.

\end{document}